\documentstyle[emulateapj,graphicx]{article}
\def\etal  {{et~al.}\ }
\def\msun{{\rm\,M_\odot}}

\def\vol#1  {{{#1}{\rm,}\ }}
\def\lya{{\rm Ly}$\alpha$\ }

\def\etal{et al.\ }

\def\gsim{\;\rlap{\lower 2.5pt
 \hbox{$\sim$}}\raise 1.5pt\hbox{$>$}\;}
\def\lsim{\;\rlap{\lower 2.5pt
   \hbox{$\sim$}}\raise 1.5pt\hbox{$<$}\;}

\newcount\refno
\newcount\rfno
\def\eq{$^{\the\refno\ }$\advance\refno by 1}
\def\ad{\advance\rfno by 1}

\def\clock{\count0=\time \divide\count0 by 60
     \count1=\count0 \multiply\count1 by -60 \advance\count1 by \time
\number\count0:\ifnum\count1<10{0\number\count1}\else\number\count1\fi}

\def\myputfigure#1#2#3#4#5%
{\hskip0.03\textwidth\vskip#5pt%\hfill
\makebox[0pt]{\hskip#2in
\includegraphics[width=#3\textwidth]{#1}}\vskip#4pt\hfill}
\newcommand{\beq}{\begin{equation}}
\newcommand{\eeq}{\end{equation}}


\begin{document}
\title{Constraining Reionization with the Evolution of the \\ Luminosity
Function of Lyman $\alpha$ Emitting Galaxies}

\author{Zolt\'{a}n Haiman}
\affil{Department of Astronomy, Columbia University, 550 West 120th Street,
New York, NY 10027} \and
%\vspace{-0.5\baselineskip}
\author{Renyue Cen} \affil{Department of Astrophysical Sciences,
Princeton University, Peyton Hall, Ivy Lane, Princeton, NJ 08544}

\begin{abstract}
At redshifts beyond $z\gsim 6$, as the mean fraction of neutral
hydrogen $\langle x_{\rm HI} \rangle$ in the intergalactic medium
(IGM) increases, the line flux of \lya emitters can be significantly
suppressed, which can result in a decrease in the observed number of
emitters at a given \lya flux.  However, cosmological HII regions
surrounding the \lya emitting galaxies alleviate these effects.  We
use simple models of the \lya line suppression that incorporate the
presence of HII regions to predict the overall effect of the \lya
absorption on the \lya luminosity function.  We find, in agreement
with other recent studies, that unless ionizing sources are unusually
strongly clustered, a fully neutral IGM may be inconsistent with the
large abundance of confirmed $z=6.5$ \lya emitters.  However, the
presence of local HII regions prohibits placing a tight constraint on
the mean neutral fraction. We find $\langle x_{\rm HI} \rangle \lsim
0.25$; the presence of strong winds and/or the clustering of ionizing
sources would further weaken this constraint. We conclude that the
evolution of the \lya LF is consistent with reionization occurring
near this redshift. Finally, we suggest that a measurement of observed
\lya line width as a function of the \lya luminosity, in a future,
larger sample of \lya emitters, may serve as a robust diagnostic of
the neutral fraction in the IGM.
\end{abstract}

\keywords{cosmology: theory -- cosmology: early universe -- galaxies:
formation -- galaxies: high-redshift -- galaxies: quasars: general --
galaxies: quasars: absorption lines}

\section{Introduction}

The recent detection of Gunn--Peterson (GP) troughs in the spectra of
the SDSS quasars (Becker et al. 2001; White et al. 2003) places a
relatively direct lower limit on the mean (volume averaged) neutral
fraction of the IGM at $z\sim 6$. While this limit is relatively weak,
$\langle x_{\rm HI} \rangle\gsim 10^{-3}$, the spectral imprint
(Mesinger \& Haiman 2004) and size (Wyithe \& Loeb 2004a) of
cosmological HII regions around the SDSS quasars indicate that the
neutral fraction around $z\sim 6$ is significantly higher, $\langle
x_{\rm HI} \rangle\gsim 0.1$.  The rapid evolution of $\langle x_{\rm
HI} \rangle$ with redshift, inferred from the spectra of several
quasars between $5.5\lsim z\lsim 6.5$ (Fan \etal\ 2002; Cen \&
McDonald 2002; Lidz et al. 2002; but see Songaila 2004), and also the
thermal state of the IGM at $3\lsim z \lsim 4$ (Hui \& Haiman 2003),
indicate and that a percolation of discrete HII regions is occurring
around this epoch.

Ly$\alpha$ emission lines from high--redshift sources can serve as
another probe of the ionization state of the IGM. The damping wing of
the GP absorption from the IGM can cause a characteristic absorption
feature (Miralda-Escud\'e 1998).  In a significantly neutral IGM, the
absorption can produce conspicuous effects, i.e. attenuating the
emission line, making it asymmetric, and shifting its apparent peak to
longer wavelengths (Haiman 2002; Santos 2004).  These effects
generally increase with $\langle x_{\rm HI} \rangle$, and it has been
suggested that a strong drop in the abundance of \lya emitters beyond
the reionization redshift can serve as a diagnostic of a nearly
neutral IGM (Haiman \& Spaans 1999).  Recently, Malhotra \& Rhoads
(2004; hereafter MR04) and Stern et al. (2004) carried out the first
application of this technique, by comparing the luminosity functions
(LFs) of \lya emitters at $z=5.7$ and $z=6.5$. The LF shows no
evolution in this range, and this has been interpreted as evidence
against percolation taking place near $z\sim 6$.

A potential caveat for this interpretation is that $z\sim 6$ galaxies
will be surrounded by their own local cosmological HII regions, which
can significantly reduce the attenuation of the \lya line flux (Cen \&
Haiman 2000; Madau \& Rees 2000).  It has been shown that this can
render \lya lines detectable even if the galaxies are embedded in a
fully neutral IGM (Haiman 2002; Santos 2004; Cen et al. 2004),
especially when the increased transmission due to the clustering of
ionizing sources is taken into account (Furlanetto et al. 2004; Gnedin
\& Prada 2004; Wyithe \& Loeb 2004b), and/or if the \lya emission has
a significant recession velocity with respect to the absorbing gas.

In this {\it Letter}, we study the impact of IGM absorption on the
\lya LF.  Our analysis is similar to that of the recent analysis by
MR04; the improvement is that we model the attenuation of Ly$\alpha$
lines including cosmological HII regions.  While very little
attenuation (by a factor of $\lsim 2$) can be tolerated by the lack of
evolution of the LF over the range $5.7\lsim z\lsim 6.5$, we find that
the presence of local HII regions allows a neutral fraction as high as
$\langle x_{\rm HI} \rangle=0.25$. On the other hand, $\langle x_{\rm
HI} \rangle\sim 1$ is allowed only if the ionizing sources are
unusually strongly clustered. In either case, we argue that the
present \lya LFs are consistent with reionization occurring near
$z\sim 6$.\footnote{Throughout this paper we assume a cold--dark
matter cosmology ($\Lambda$CDM), with ($\Omega_\Lambda$, $\Omega_M$,
$\Omega_b$, $H_0$) = (0.73, 0.27, 0.044, 71 km s$^{-1}$ Mpc$^{-1}$),
consistent with the recent results from {\it WMAP} (Spergel et
al. 2003).  Unless stated otherwise, all lengths are quoted in
comoving units.}

\section{The Observed Ly$\alpha$ Luminosity Function}

Following Malhotra \& Rhoads (2004; hereafter MR04) we take as basic
inputs to our analysis the \lya LFs at $z=5.7$ and $z=6.5$.  We adopt
the forms of the LF at these two redshifts from MR04.  Note that at
both redshifts, the LFs are inferred by culling several independent
datasets (Hu et al. 2002; Rhoads et al. 2003; Kodaira et al. 2004).
At $z=5.7$, the LF is comparatively well determined and well fit by a
Schechter function with relatively small errors in normalization
($\Phi_\star$) and characteristic luminosity ($L_\star$), while the
faint--end slope ($\alpha$) is less well--determined.  Here we adopt
the best--fit values of $(\Phi_\star,L_\star,\alpha) = (10^{-4}~{\rm
Mpc^{-3}},10^{43}~{\rm erg~s^{-1}}, -1.5)$ (displayed by the
short--dashed curve in Fig.~\ref{fig:LFs} below). Within the
uncertainties quoted in MR04, the values of these parameters do not
significantly change our conclusions below.

\myputfigure{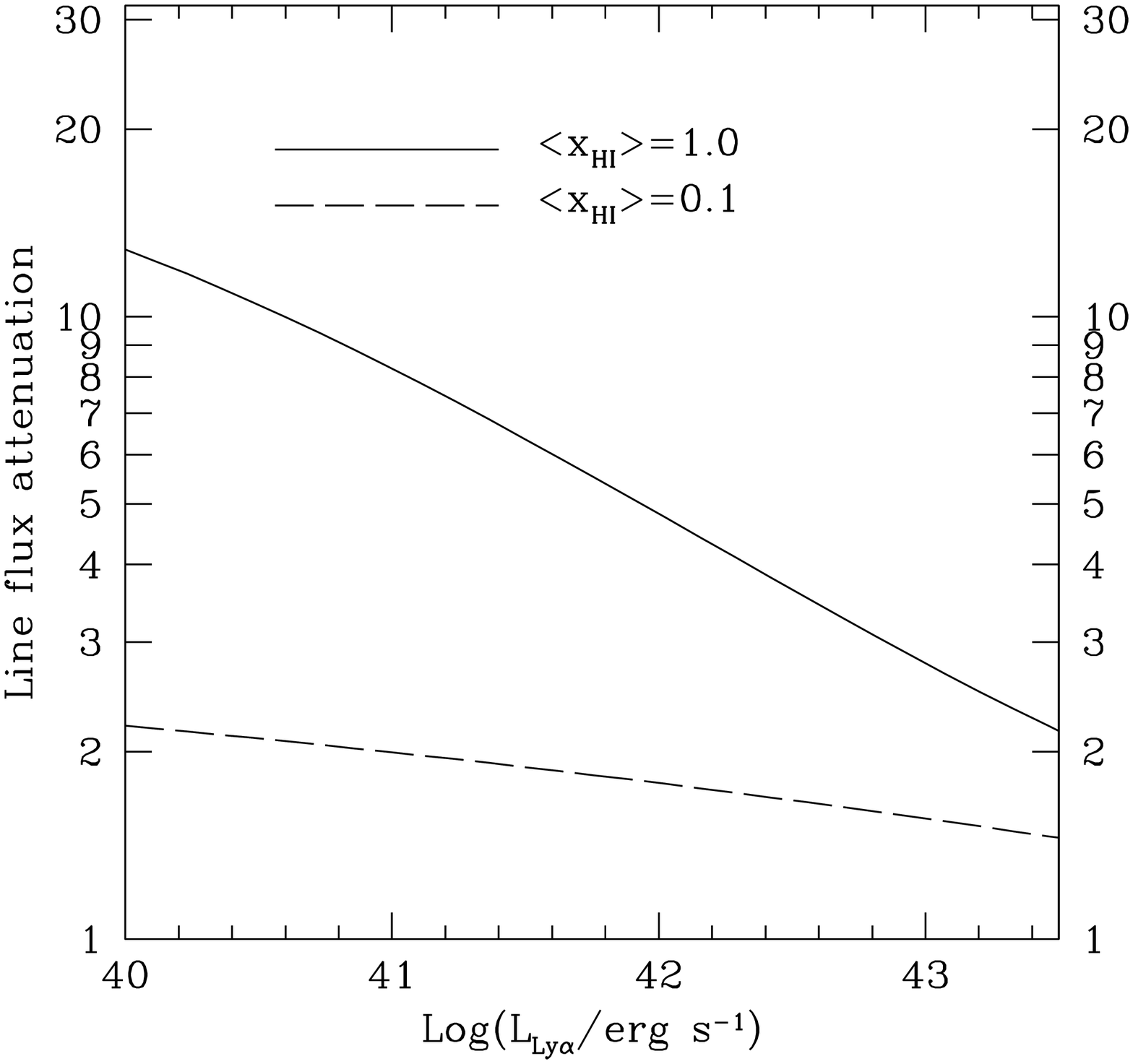}{3.2}{0.45}{-10}{-10} \figcaption{The
suppression of the total Ly$\alpha$ line flux, as a function of the
observed line luminosity, for an IGM with $\langle x_{\rm HI} \rangle
= 0.1$ (or $\Gamma_{12}=0.0015$; long--dashed curve), and for and IGM
with $\langle x_{\rm HI} \rangle = 1.0$ (or $\Gamma_{12}=0$; solid
curve).
\label{fig:suppression}}
\vspace{\baselineskip}

At $z=6.5$, the LF is less well determined, as shown by four
data--points with errors in Figure~\ref{fig:LFs} below.  Of these four
data--points, the most constraining is from the deep Subaru field,
which revealed the presence of a handful of emitters at $z\sim 6.5$
(Kodaira et al. 2003), and has the smallest error bar.  The single,
faint lensed galaxy at this redshift found by Hu et al. (2002;
represented by the lowest--luminosity point in Fig.~\ref{fig:LFs}
below) implies a surprisingly high abundance of the faintest emitters,
but with a single source, it has a large uncertainty.  Finally, we
follow MR04 and conservatively assume that the intrinsic LF
(i.e. prior to processing of the line through the IGM) is the same at
$z=6.5$ as at $z=5.7$.  More realistic models, based on the
hierarchical growth of structures, combined with \lya line processing
by gas and dust interior to the galaxies, predict that the number of
emitters were smaller at earlier epochs (Haiman \& Spaans 1999; Le
Delliou et al. 2004), which would strengthen the limits on $\langle
x_{\rm HI}\rangle$ below.

\section{The Attenuation of \lya Emission Lines}
\label{sec:transmission}

We follow Cen \& Haiman (2000), and use a simple model to find the
attenuation of the Ly$\alpha$ line as a function of wavelength.  The
model is straightforward, and we only briefly recap the main
features. We start with a Gaussian emission line originating at
$z_s=6.5$, with a line--width of $\Delta v=300~{\rm km~s^{-1}}$,
typical of normal galaxies (narrower lines would be attenuated more
severely; see discussion below). We assume that the source is
surrounded by a spherical Str\"omgren sphere that propagated into the
IGM with a neutral fraction $\langle x_{\rm HI}\rangle$.  The IGM is
assumed to have a clumping factor
$C\equiv\langle\rho^2\rangle/\langle\rho\rangle^2=10$ with a
log-normal distribution.  We assume the presence of a uniform ionizing
background, characterized by an ionizing rate $\Gamma_{12}$
(ionizations per $10^{12}$s per atom, yielding the given value of
$\langle x_{\rm HI}\rangle$ in ionization equilibrium for the IGM.

We assume a Salpeter IMF, and relation (Kennicutt 1983) between
star--formation rate and \lya luminosity (SFR=$1 \msun$/yr)
$\rightarrow$ ($L({\rm Ly}\alpha)=10^{42}$ erg/s). We conservatively
assume $f_{\rm esc}$ = 1 (escape fraction of ionizing radiation) in
computing $R_s$, and a lifetime $t_s=10^8$ yrs for each source.  With
all other parameters in the model fixed, we vary $\Gamma_{12}$ (or
equivalently $\langle x_{\rm HI} \rangle$), and compute the factor by
which the total line flux is suppressed relative to the unabsorbed
Gaussian line. This factor is shown in Figure~\ref{fig:suppression} as
a function of the intrinsic luminosity of the source. Faint sources
are more severely attenuated: by a factor of up to $\sim 15$ for $\log
L($\lya$/{\rm erg~s^{-1}})=40$ [or up to $\sim 6$ for $\log
L($\lya$/{\rm erg~s^{-1}})=42$, the present lower limit for the
detection of a \lya emitter at $z=6.5$] in a neutral universe.  Bright
($\log L($\lya$/{\rm erg~s^{-1}})=43$) sources, however, are only
attenuated by a factor of $\sim 1.5$ even if $\langle x_{\rm HI}
\rangle$ is as high as 0.1.

The attenuation factors in Figure~\ref{fig:suppression} are computed
by assuming that the HII regions are being created by a single
ionizing source (the \lya emitter itself). These HII regions (with
typical sizes of a few Mpc in the observed range of source
luminosities) may contain many other undetected galaxies, which make
the HII regions larger (Furlanetto et al. 2004; Gnedin \& Prada 2004;
Wyithe \& Loeb 2004b) and also more highly ionized.  Therefore, the
results shown in Figure~\ref{fig:suppression} should be viewed as
conservative upper limits.  These results show that the HII regions of
individual galaxies can significantly reduce the attenuation,
especially for relatively bright sources.

\section{The Impact on the Ly$\alpha$ Luminosity Function}

We next use the attenuation factors computed above, and obtain the
reduction in the cumulative \lya LF.  The resulting LF is shown in
Figure~\ref{fig:LFs}, for three values of $\langle x_{\rm HI}
\rangle$: 0, 0.1, and 1 (top to bottom).  The plot also shows the data
for the $z=6.5$ LF with error bars (from MR04).  The short--dashed
curve is a reproduction of the $z=5.7$ LF (a Schechter function),
which, by assumption, would be identical to the $z=6.5$ LF if the IGM
contained no neutral HI.

Note that the suppression of the line flux depends on the luminosity
of the source, but the reduction of the LF in Figure~\ref{fig:LFs} is
uniform, by a nearly constant factor. The steepening of the LFs at the
bright end compensates for the reduced effect of line attenuation; the
two conspire to give a nearly constant suppression of the cumulative
LF.

\myputfigure{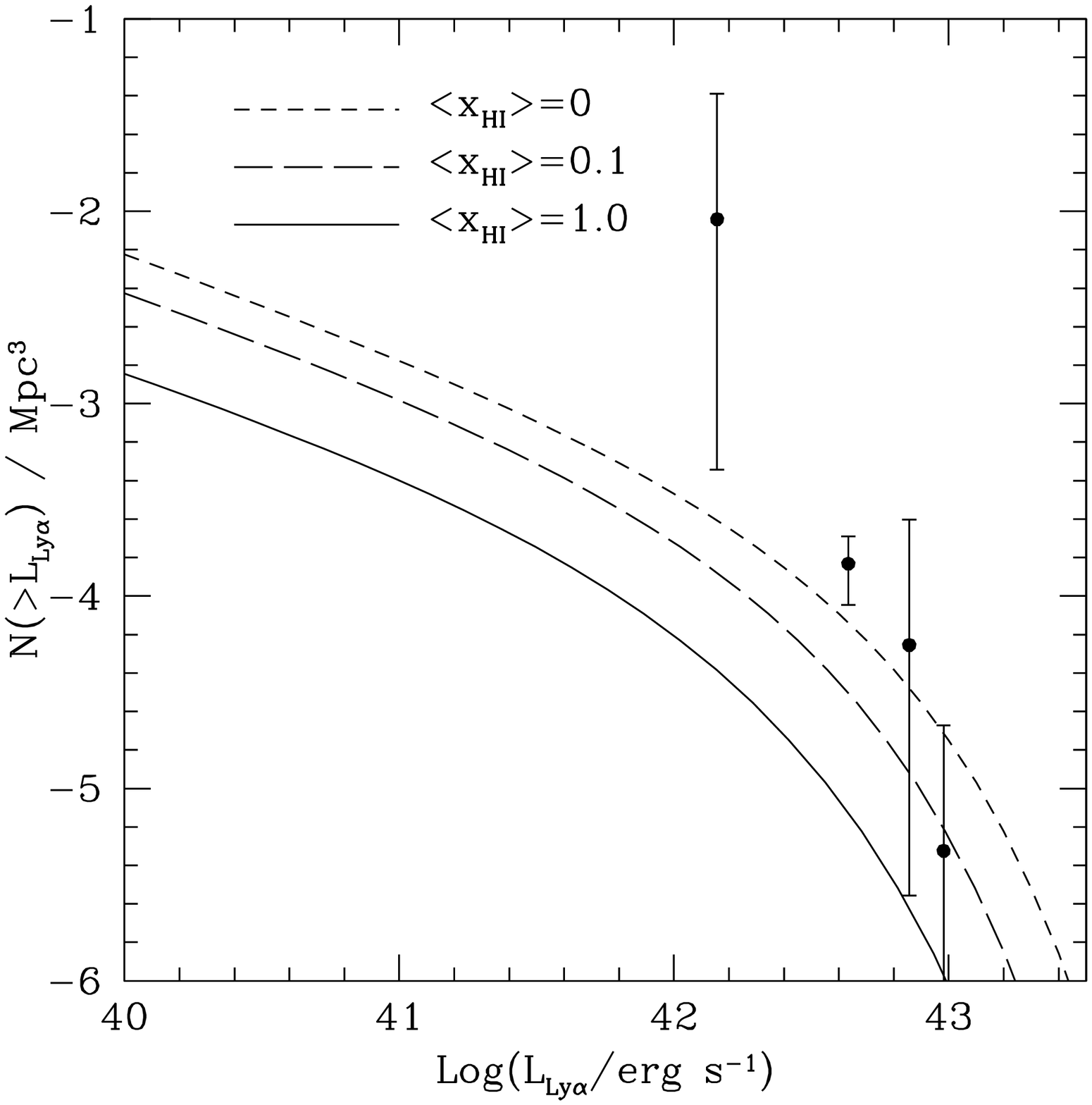}{3.2}{0.45}{-10}{-10}
\figcaption{The Ly$\alpha$ luminosity functions for a
fully ionized IGM (no suppression, short--dashed curve), for an IGM with
$\langle x_{\rm HI} \rangle = 0.1$ (or $\Gamma_{12}=0.0015$;
long--dashed curve), and for and IGM with $\langle x_{\rm HI} \rangle = 1.0$
(or $\Gamma_{12}=0$; solid curve).
\label{fig:LFs}}
\vspace{\baselineskip}

\section{Results and Discussion}

We can now assign a likelihood to each value of $\langle x_{\rm
HI}\rangle$, by comparing the model LFs and the data shown in
Figure~\ref{fig:LFs} . For each value of $\langle x_{\rm HI} \rangle$,
we obtain the mean expected number of \lya emitters above the flux
thresholds shown by the data points (here we use the effective volume
probed by each survey, listed in Table 2 in MR04).  We assume Poisson
fluctuations to compute the likelihood in each bin.  We find that the
lowest--luminosity data--point has an a--priori Poisson likelihood
that is exceedingly low, even in a fully ionized universe. As
Figure~\ref{fig:LFs} shows, $\lsim 0.03$ objects are expected in the
small survey volume in which the single lensed \lya emitter,
represented by this data--point, was discovered. However, given the
difficulties in obtaining an accurate effective volume from the poorly
constraint lensing configuration for this object, we follow MR04 and
omit this source from our analysis.  The result, i.e. the product of
the Poisson probabilities for the three higher luminosity bins, are
shown in Figure~\ref{fig:probs} as a function of $\langle x_{\rm HI}
\rangle$.

We find that $\langle x_{\rm HI}\rangle=1$ is ruled out at high
significance, in agreement with the conclusions reached by MR04 and
Stern et al. (2004) (but see discussion below).  On the other hand, at
the 99.7\% confidence level, we find only the relatively weak
constraint $\langle x_{\rm HI} \rangle < 0.25$.  The upper dashed line
corresponds to the limit of no HI absorption and shows that the data
is consistent (at the $\sim 10\%$ confidence level) with the intrinsic
$z=5.7$ and $z=6.5$ LFs being identical (but note that the IGM would
have to be ionized to an unrealistically high level to actually
approach this limit).

\myputfigure{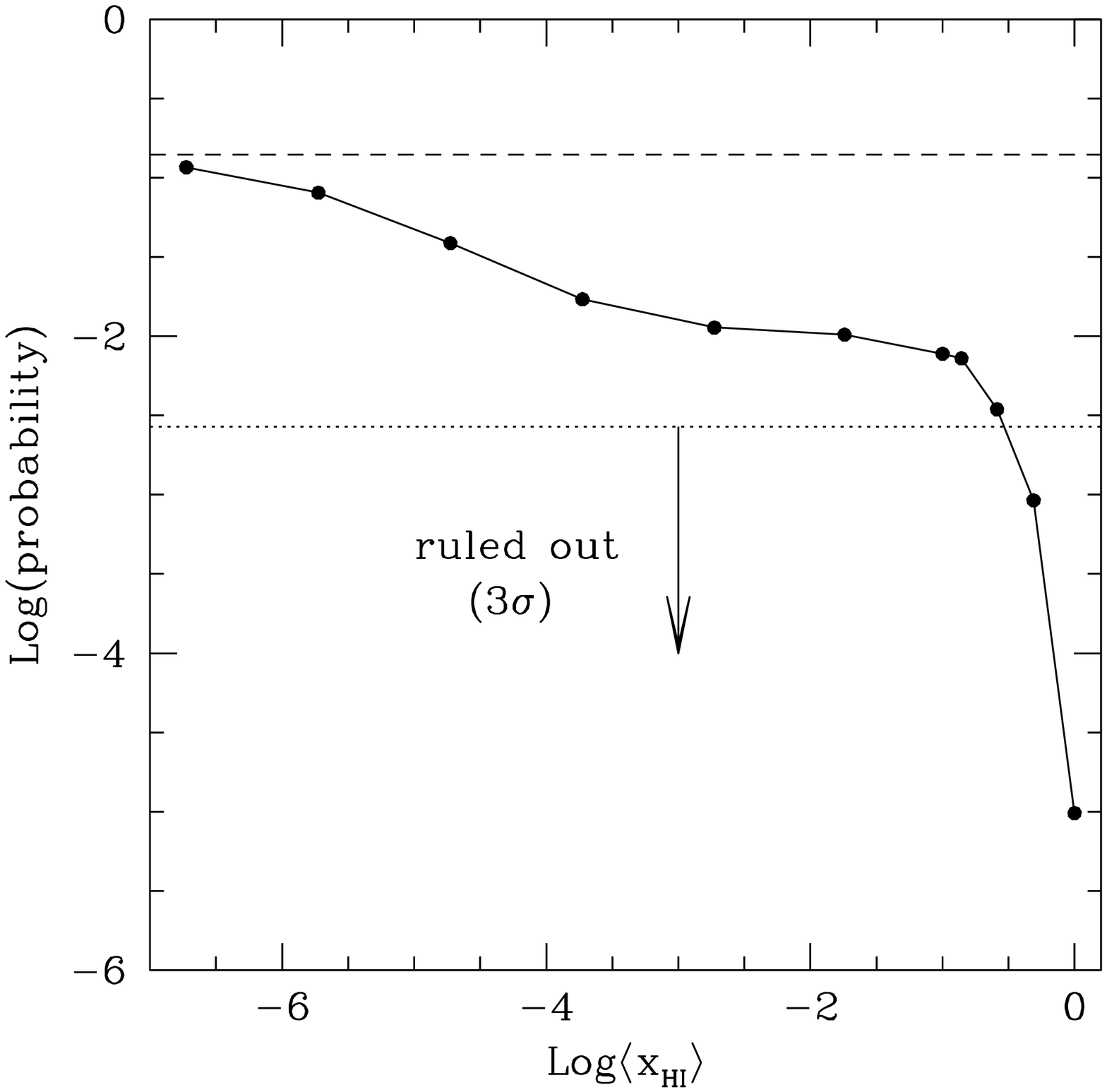}{3.2}{0.45}{-10}{-10} \figcaption{The
probability of each mean neutral fraction in the IGM, given the
observational constraints from the $z=6.5$ Ly$\alpha$ luminosity
function.  The model assumes that the intrinsic LF is the same as at
$z=5.7$ (see text for discussion).\label{fig:probs}}
\vspace{\baselineskip}

As mentioned above, HII regions can be enlarged due to the clustering
of ionizing sources, and also due to their being elongated along our
line of sight (e.g. Gnedin \& Prada 2004).  These effects reduce the
\lya attenuation, and weaken the upper limit on $\langle x_{\rm
HI}\rangle$.  It is interesting to ask whether {\em any} constraint
can be placed on the neutral fraction, in the presence of
clustering. To address this question, we re--computed the attenuation
factors shown in Figure~\ref{fig:suppression}, and the implied LFs
shown in Figure~\ref{fig:LFs}, fixing $\langle x_{\rm HI}\rangle=1$,
but boosting the ionizing emissivity of each source by a constant
factor of $B$.  We find that the boosting factor required to render
this model acceptable at the 99.7\% confidence level is approximately
$B=60$ (note that the increased luminosity is required to reduce both
the GP damping wing, and the residual HI inside the HII region).

In order to assess whether the ionizing emissivity can be boosted by
this factor, we computed the total mass in dark matter halos in the
mass range $M_{\rm min}$ to $M_{\rm s}$ and within a sphere of radius
$R$ around a halo of mass $M_{\rm s}$. We included the spatial
correlations between halos with a prescription for linear bias (see
eq. 15 in Haiman et al. 2001).  For the likely halo mass range of
$10^{11}{\rm M_\odot}\lsim M \lsim 10^{12}{\rm M_\odot}$ (Rhoads et
al. 2003) and HII sphere radii ($2\lsim R \lsim 6$ Mpc; e.g. Haiman
2002) of \lya emitters at $z=6.5$, we find that the total collapsed
halo mass, down to $2\times 10^{8}{\rm M_\odot}$ (corresponding to a
virial temperature of $10^4$K at $z=6.5$) is increased by a factor of
$2-18$ (the satellite halos have linear bias parameters of $2\lsim
b\lsim 10$). Unless the star formation efficiency is preferentially
higher in the low--mass ``satellite'' galaxies, it is unlikely that
the ionizing emissivity is boosted by the required factor of $60$.  A
recent numerical simulation by Gnedin \& Prada (2004, see their
fig. 3) however, suggests that at least $10\%$ of galaxies may
transmit their \lya lines without any absorption at $z=6.5$.

\myputfigure{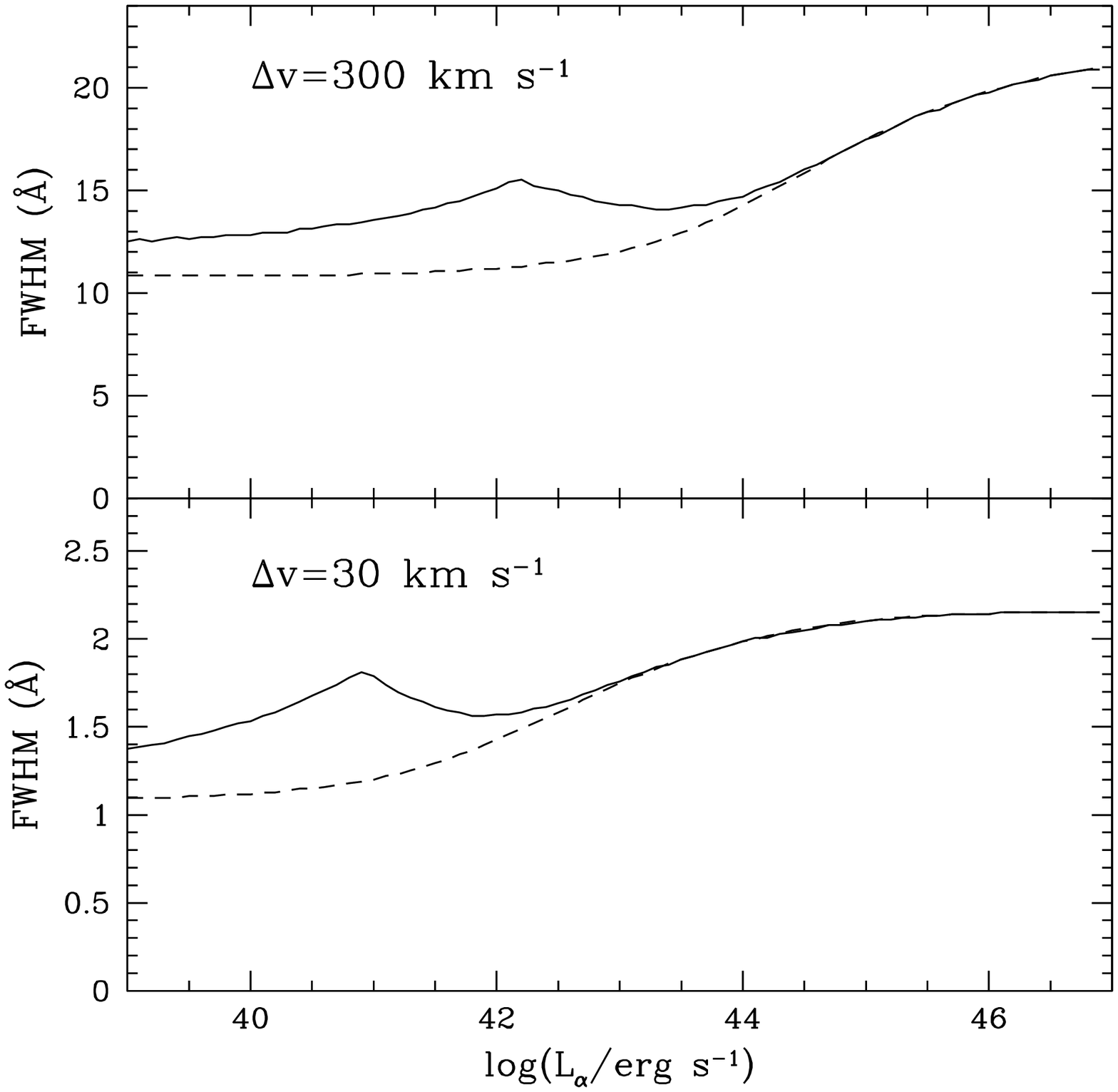}{3.2}{0.45}{-10}{-10} \figcaption{The FWHM of
the transmitted \lya line as a function of the luminosity of the
source.  The upper (lower) panel assumes an intrinsic Gaussian line
shape with a width of 300 (30) km/s.  On both panels, the dashed curve
assumes an ionizing background of $\Gamma_{12}=0.1$ (or $\langle
x_{\rm HI} \rangle\sim 10^{-3}$), and the solid curve assumes
$\Gamma_{12}=0$ ($\langle x_{\rm HI} \rangle = 1$).  The feature at
$\sim 10^{41-42}~{\rm erg~s^{-1}}$ is unique to a significantly
neutral IGM.
\label{fig:fwhm}}
\vspace{\baselineskip}

Another effect that can further weaken the upper limits on the neutral
fraction is a systematic redshift of the \lya emission lines relative
to the absorbing gas.  Lyman break galaxies at lower redshift
reveal such systematic shift by several hundred km/s, attributed to
galactic winds (Shapley et al. 2003).  If velocity offsets as high as
this are common for \lya emitters at $z\sim 6$, this would render \lya
line attenuation from the residual HI within the HII region
effectively negligible, and will also significantly reduce the GP
damping wing absorption.  By repeating our analysis above to obtain a
likelihood for a neutral universe, we find that a redshift by $600$
km/s is required to for $\langle x_{\rm HI} \rangle = 1$ to yield an
acceptable fit for the LF.  Shapley et al. (2003) find a decreasing
velocity offset with increasing \lya equivalent width, with $\langle
\Delta v\rangle<500$ km/s for $W_\alpha>50$\AA\ (see their fig.~11),
suggesting that high--$z$ \lya emitters may not have the requisite
$600$ km/s offsets. 

Finally, we propose a different diagnostic of the neutral fraction
that could be available from a future, larger sample of $z\gsim 6$
\lya emitters.  For faint \lya emitters, the line attenuation is
dominated by the residual HI inside their cosmic HII regions, whereas
for bright emitters, the GP damping wing is dominant (Haiman 2002).
In Figure~\ref{fig:fwhm}, we show the predicted full width at half
maximum (FWHM) of the transmitted line, as a function of \lya
luminosity (for two different intrinsic line--widths). The figure
reveals a clear imprint of this transition at $\sim 10^{41-42}~{\rm
erg~s^{-1}}$.  This feature occurs only if the IGM contains a
significant amount of neutral hydrogen (solid curves), and we checked
that it is present for different assumed intrinsic line--shapes
(e.g. a top--hat or a Lorentzian, which have less/more extended wings
than a Gaussian). This suggests that a measurement of observed \lya
line width as a function of the \lya luminosity may therefore serve as
a more robust diagnostic of the neutral fraction.

\section{Conclusions}

In this {\it Letter}, we have considered the attenuation of the line
flux of \lya emitting galaxies at $z=6.5$.  We included the effect of
HII regions around these galaxies, which reduce the line
attenuation. We computed the overall impact of the IGM on the \lya
luminosity function, and found that the presence of the HII regions
makes a large mean neutral fraction in the IGM acceptable. The best
upper limit is at most as strong as $\langle x_{\rm HI} \rangle \lsim
0.25$. However, a fully neutral IGM may be acceptable only in the
presence of an unusually strong clustering of ionizing sources, and/or
a strong systematic recession velocity of the \lya emitting gas.  In
either case, the present determination of the $z\sim 6.5$ \lya LF is
consistent with reionization occurring at this epoch. The LF of a
larger sample of \lya emitters, or a systematic study of their
linewidths and shapes, as a function of luminosity, should
nevertheless provide a robust diagnostic of the ionization state of
the IGM in the future.

\acknowledgments{This work was supported by NASA through grants
NAG5-26029, HST-GO-09793.18 (ZH) and NAG5-13381 (RC), and by NSF
through grants AST-0307200, AST-0307291 (ZH), and AST-0206299,
AST-0407176 (RC).}

\end{document}